\documentclass[twocolumn]{aastex62}

\usepackage{threeparttable}
\graphicspath{{./}{figures/}}

\received{\today}
\revised{\today}

\submitjournal{ApJ}

\shorttitle{Genus Statistic of Tycho's Supernova Remnant}
\shortauthors{Sato et al.}

\begin{document}

\title{Genus Statistic Applied to the X-ray Remnant of SN 1572: Clues to the Clumpy Ejecta Structure of Type Ia Supernovae}

\correspondingauthor{Toshiki Sato}
\email{toshiki.sato@riken.jp}

\author[0000-0001-9267-1693]{Toshiki Sato}
\affil{RIKEN, 2-1 Hirosawa, Wako, Saitama 351-0198, Japan}
\affil{NASA, Goddard Space Flight Center, 8800 Greenbelt Road, Greenbelt, MD 20771, USA}
\affil{Department of Physics, University of Maryland Baltimore County, 1000 Hilltop Circle, Baltimore, MD 21250, USA}

\author[0000-0002-8816-6800]{John P.Hughes}
\affiliation{Department of Physics and Astronomy, 
Rutgers University, 136 Frelinghuysen Road, Piscataway, 
NJ 08854-8019, USA}

\author[0000-0003-2063-381X]{Brian J. Williams}
\affiliation{NASA, Goddard Space Flight Center, 8800 Greenbelt Road, Greenbelt, MD 20771, USA}

\author{Mikio Morii}
\affiliation{The Institute of Statistical Mathematics, 
10-3 Midori-cho, Tachikawa, Tokyo 190-8562, Japan}



\begin{abstract}
Clumpy structures are a common feature in X-ray images of young Type Ia supernova remnants (SNRs).  Although the precise origin of such clumps remains unclear there are three generic possibilities: clumpiness imposed during the explosion, hydrodynamic instabilities that act during the remnant's evolution, and pre-existing structures in the ambient medium. In this article we focus on discriminating between clumping distributions that arise from the explosion and those from the remnant's evolution using existing 3D hydrodynamical simulations.  We utilize the genus statistic for this discrimination, applying it to the simulations and {\it Chandra}  X-ray observations of the well-known SN Ia remnant of SN 1572 (Tycho's SNR).  The genus curve of Tycho's SNR strongly indicates a skewed non-Gaussian distribution of the ejecta clumps and is similar to the genus curve for the simulation with initially clumped ejecta.  In contrast, the simulation of perfectly smooth ejecta where clumping arises from the action of hydrodynamic instabilities produced a genus curve that is similar to a random Gaussian field, but disagrees strongly with the genus curve of the observed image. Our results support a scenario in which the observed structure of SN Ia remnants arises from initial clumpiness in the explosion.


\end{abstract}



\section{Introduction} \label{sec:intro}
The X-ray emission from young supernova remnants (SNRs) holds important clues about the nature of supernova explosions.  This is particularly relevant for the remnants of Type Ia supernovae (SNe Ia) which are used as standardizable candles to determine the expansion history of the universe \citep[e.g.,][]{1998AJ....116.1009R,1999ApJ...517..565P}. 
In particular, since the ejecta of SN Ia remnants emit primarily in the X-ray band as optically thin thermal emission and are extended across many arcminutes, the X-ray morphology of such remnants can provide unique information about the spatial structure of the explosions. 

An obvious, but as yet poorly quantified, feature of SN Ia remnants is the clumpy appearance of their X-ray--emitting ejecta. Although first noted in X-ray images of young SNRs by the {\it Einstein} observatory in the early 1980's \citep[such as Tycho's SNR:][]{1983ApJ...266..287S}, the processes by which the ejecta structures form remain obscure despite intense observational and theoretical efforts. 

Structure in a remnant is the consequence of initial conditions (i.e., structure imposed during the explosion), evolutionary effects (e.g., hydrodynamic instabilities), and any pre-existing structure in the ambient medium.  
Since the contact surface between the ejecta and the ambient medium is subject to Rayleigh-Taylor and Kelvin-Helmholtz dynamical instabilities \citep{1975MNRAS.171..263G}, the ejecta in a young remnant will appear as clumpy, even in the absence of structure in the initial ejecta distribution and the ambient medium.  
In one study focused on SN Ia remnants
\cite{2001ApJ...549.1119W} used two-dimensional hydrodynamical simulations to study the structure generated
by the dynamical instabilities and also by introducing overdense ejecta clumps of varying size. 
They found that ejecta clumps could survive as distinct features for several hundred years as long as the clump's initial overdensity  (compared to the surrounding ejecta) was high enough (at least a factor of 100 in the
case of Tycho's SNR). The magnetohydrodynamic simulations of \cite{2012ApJ...749..156O} showed that magnetic-field amplification at the outer edges of clumps could stabilize them, enabling clumps with a smaller density contrast (a factor of only 10) to survive. Both of these previous studies introduced ejecta clumps as {\it ad hoc} features in an otherwise smooth ejecta distribution.  \cite{2013MNRAS.429.3099W} ran high-resolution three-dimensional (3D) hydrodynamic simulations of SN Ia remnants assuming smooth (unclumped) ejecta interacting with a uniform interstellar medium, while varying the compressibility of the gas to approximate the effect of efficient particle acceleration \citep[e.g.,][]{2000ApJ...543L..57D}.  These authors argued that the dynamical instability was sufficient to describe the radial profile and clumpy central structure of SN Ia remnants and that clumpiness in the initial ejecta distribution was not required.



Tycho's SNR, studied by Tycho Brahe in 1572, is known to be the remnant of a normal SN Ia based on its light-echo spectrum \citep{2008Natur.456..617K}.  As noted above, clumpy ejecta structures in the remnant were clearly recognized in early broad-band high resolution ($\sim$5$^{\prime\prime}$) X-ray imaging with the {\it Einstein} and {\it ROSAT} missions \citep[e.g.,][]{1983ApJ...266..287S,2000ApJ...545L..53H}.  These data also showed evidence for spectral inhomogeneities in the form of two clumps along the rim in the southeastern quadrant: one Si-rich and the other Fe-rich \citep{1995ApJ...441..680V}, which were subsequently confirmed by follow-up studies with X-ray CCD observations \citep{1997ApJ...475..665H,2001A&A...365L.218D}. High resolution ($\sim$1$^{\prime\prime}$) imaging with {\it Chandra} \citep{2002ApJ...581.1101H,2005ApJ...634..376W} revealed that the structure of the ejecta emission was quite clumpy on scales of a few arcseconds, but without the compact knots seen in the core collapse SNR Cas A \citep{2000ApJ...528L.109H,2000ApJ...537L.119H}.

Recently, \cite{2017ApJ...840..112S} found that X-ray clumps in Tycho's SNR  could be separated into distinct redshifted and blueshifted ejecta clumps using {\it Chandra} imaging spectroscopy, giving researchers new insights into the three-dimensional kinematics of Type Ia SN ejecta.  In addition, \cite{2017ApJ...842...28W} compared the ejecta kinematics in Tycho's SNR (specifically the range of deceleration parameters from proper motion measurements of individual clumps) with a pair of 3D hydrodynamical models of the remnant's evolution: one model started with a smooth initial ejecta distribution and the other started with a clumped ejecta distribution.  These authors argued that both models, when evolved to the age of Tycho's SNR, were able to accommodate the observed distribution of deceleration parameters for the ejecta. On the other hand, the simulated images were distinctly different in visual appearance, suggesting that the current morphology of the remnant contains usable information on the initial distribution of the ejecta. 

Our preliminary investigations into constraining the structure of SN Ia remnants using Fourier and wavelet-transform \citep[e.g.,][]{2009ApJ...691..875L} analyses did not turn out to be sufficiently powerful at discriminating the two hydro models and the observed Tycho image from each other.  This led us to investigate an approach that would be more sensitive to patterns in the distribution of clumps and holes in the images, such as the "genus statistic."
The genus statistic in astronomy was initially used to study the topology of the universe and the 3D distribution of galaxies \citep{1986ApJ...306..341G,1987ApJ...319....1G}. 
It was also applied to \ion{H}{1} column density maps of the Small Magellanic Cloud \citep{2008ApJ...688.1021C} and density fluctuations in MHD simulations of turbulence \citep{2007ApJ...658..423K}. Our study represents the first time the genus statistic has been applied to SNRs. 

In the following we present a brief summary of the genus statistic (\S 2). Then in section 3 we report on our determination of the genus curve from the X-ray images of Tycho's SNR, addressing issues of image smoothing, variations among different observations, our methods to determine uncertainties on the genus curve, and the variations of the genus curves for different observed energy bands.  In section 4 we evaluate the genus curves for the simulations and compare them to the data.  Here we also discuss the implications of our work in the context of explosion models for SN Ia.  We conclude with a summary of results in section 5.  Throughout this paper, unless explicitly stated, we quote uncertainty intervals at the 90\% confidence level.

\section{Genus Statistic} \label{sec:Genus}
\citet{1988MNRAS.234..509C} studied the statistical geometry of 2D random fields. He began with the idea of an excursion set (the set of points in an image above a given intensity threshold) and showed how the Euler-Poincar\'e (EP) characteristic\footnote{The EP characteristic is the integral over all contours in the map (at a given intensity level) of the geodesic curvature of the contour \citep[for more details see][]{1988MNRAS.234..509C}.} of the excursion set could provide a statistic sensitive to the pattern of holes and peaks in 2D images generated with different types of random fields. 
The EP characteristic of a 2D image can be represented as 
\begin{equation} 
        G(I_0) \equiv N_{>I_0} - N_{<I_0}
\end{equation}
where $N_{>I_0}$ is the number of isolated regions in the image data above a threshold intensity $I_0$ and $N_{<I_0}$ is the number of regions below the threshold. Hereafter we refer to $G$ as the genus (although we are aware that this nomenclature is not,  strictly speaking, mathematically correct). The genus curve of a 2D image is constructed by varying the threshold $I_0$ over the range of intensities in the data and typically the genus is expressed as a function of $\nu$ where $I_0 = \nu \sigma$ and $\sigma$ is the root-mean-square of the map intensity.  Here, a filled circle would produce a genus of 0, while a ring shape that has one ``hole" would be classified with a genus of $-1$. Furthermore, 2 (or 3, 4...) disjoint filled circles would be classified as genus of 1 (or 2, 3...).  

\begin{figure}[h]
 \begin{center}
  \includegraphics[bb=0 0 835 724,width=8cm]{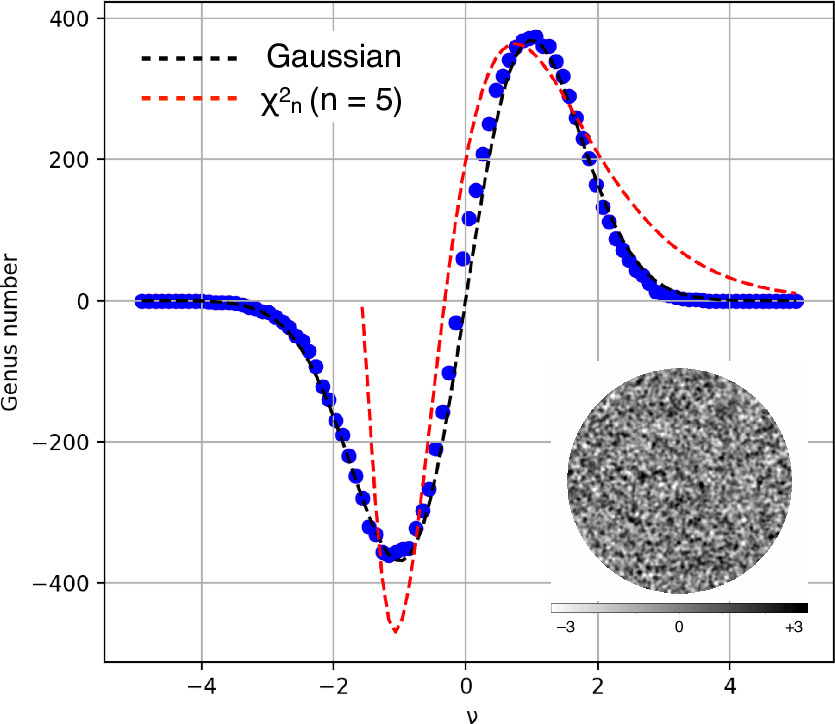}
 \end{center}
\caption{An example of the genus curve (solid blue dots) from a random Gaussian distribution, whose image is shown in the insert at lower right. The black (red) dashed line shows the best-fit genus curve using the analytic expression for the random Gaussian (chi-square pdf) field. The chi-square pdf case has $n=5$ and the same coherence angle as that for the Gaussian field.  }
\label{fig:GaussGenuss}
\end{figure}

We use the ``genus'' python module in ``TurbuStat"\footnote{https://github.com/Astroua/TurbuStat} 
\citep{2017MNRAS.471.1506K} to obtain the genus curves of our images.  Figure \ref{fig:GaussGenuss} shows an example of the genus curve obtained from a random Gaussian field. At thresholds below the mean intensity, the genus curve shows a deep dip because of a large number of low-intensity (``hole'') regions. On the other hand, at higher threshold levels, the number of high-intensity (``clump'') regions dominate. At increasingly higher or lower intensity values there are fewer  and fewer structures as the image runs out of  pixels with intensity values far from the mean. The 
genus curve for a random Gaussian distribution can be analytically expressed as \citep{1988MNRAS.234..509C}
\begin{equation} \label{GaussEq}
        G(\nu) = \frac{A}{(2 \pi)^{3/2}~\theta_{\rm c}^{2}} ~ \nu ~{\rm exp} \left( - \frac{\nu^2}{2}\right),
\end{equation}
where $A$ and $\theta_c$ are the area of the field and the coherence angle, respectively. 

Some analytic solutions for non-Gaussian fields are also given by \cite{1988MNRAS.234..509C}. Here, we introduce the analytic solution for the chi-square\footnote{We spell out the phrase ``chi-square'' when describing this analytic model for the random field and use $\chi^2$ to describe the figure-of-merit function that we use for fitting the genus curves later in the paper.} probability density function (pdf) of order $n$, which is described as
\begin{equation} \label{chiEq}
        G(u) = \frac{A}{2 ~ \Gamma(n/2)(2 \pi)^{3/2}~\theta_{\rm c}^{2}}~ [u-(n-1)] ~{\rm exp} \left( - \frac{u}{2}\right),
\end{equation}
where $\Gamma$ is the Gamma function. To convert the threshold $u$ to standardized form, $u = \mu + \nu \sigma$,  we use the elementary results from probability theory that $\mu = n$ and $\sigma^2 = 2 n$ for a chi-square pdf. The genus curve here is asymmetric in amplitude (there are typically more holes than peaks) and the $G=0$ value is offset from zero intensity toward negative values.  The genus curve of the chi-square pdf approaches that of the Gaussian distribution as $n$ grows large.

Although there is no expectation that these simple models will accurately describe the quantitative behaviour of actual X-ray images of SNRs, they are useful as heuristic guides and as a baseline for comparison between the images of Tycho's SNR and the hydrodynamical models. 

\section{Observation, Data Analysis and Results} \label{sec:obs}
\begin{figure*}[t]
 \begin{center}
  \includegraphics[bb=0 0 825 774,width=15cm]{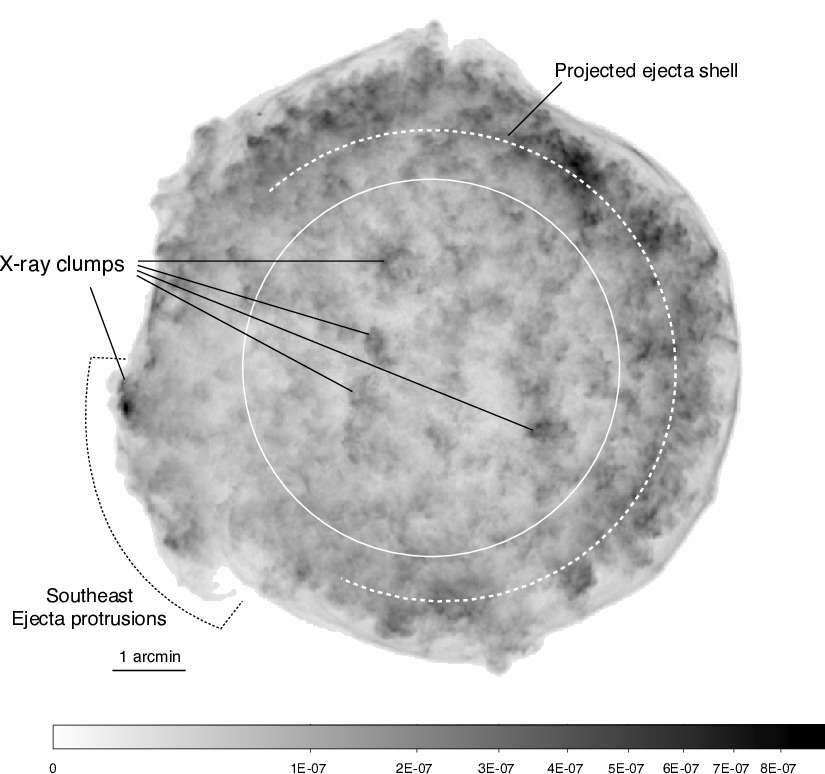}
 \end{center}
\caption{The flux image (1.76--4.2 keV) of Tycho's SNR observed in 2009 by {\it Chandra}. The white dashed line shows the location of the projected shell structure discussed in the text. The solid white circle shows the region of the remnant used in our genus statistic analysis.
}
\label{fig:img}
\end{figure*}
We aim to evaluate the clumpy structures in Tycho's SNR using {\it Chandra} high angular resolution ($\sim1^{\prime\prime}$) images. The
{\it Chandra} ACIS-I detector imaged Tycho's SNR in 2009 for an effective
exposure of 734.1 ksec (PI: John P.\ Hughes).   
We also use the shorter ACIS-I observations ($\sim 150$ ks) of the remnant taken in 2007 (PI: John P.\ Hughes) and 2015 (PI: Brian J.\ Williams) for the comparison presented in section \ref{sec:exp}.
We reprocessed all the level-1 event data, applying the standard data reduction with CALDB version 4.6.1. During this process, we use a custom pipeline based on ``\verb"chandra_repro"" in CIAO version 4.7. 

Figure \ref{fig:img} shows the flux image (including line emission from Si, S, Ar and Ca) of Tycho's SNR. In the central part of the remnant, we can clearly see the clumpy structures that we focus on in this study. At the outer radius of $\sim$3.4 arcmin, there is an intensity peak because of the projected (limb-brightened) shell structure \citep[e.g.,][]{2017ApJ...840..112S} that is composed of a large number of ejecta knots projected on top of each other. In addition, the peculiar ejecta features mentioned in the introduction appear at the edge of the southeastern rim. These structures as well as 
limb brightening complicate the evaluation of the clumpy structures. Therefore, we only use the central region 
(white circle in Figure \ref{fig:img}) where the main effect of projection comes from the central (and therefore more nearly face-on) front and back hemispheres of the expanding ejecta shell.

X-ray intensity is roughly proportional to the square of plasma density. Therefore, we expect the square root of the intensity to reflect more accurately the density distribution, and we use the image scaled in this way for the genus analysis. In order to compare directly with the theoretical images (as shown in the following section), we further scaled the images by subtracting the mean image intensity value from each pixel
and then dividing by the square root of the variance of the image intensity values.

\subsection{Smoothing and Monte Carlo Sampling} \label{sec:smoothing}
\begin{figure*}[t]
 \begin{center}
  \includegraphics[bb=0 0 929 354,width=16cm]{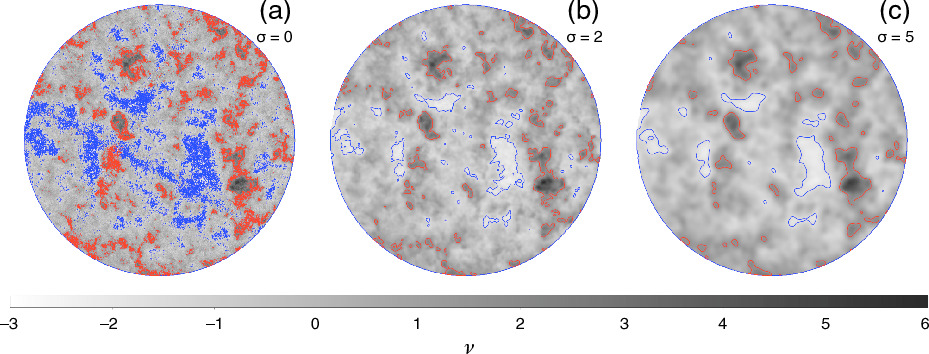}
 \end{center}
\caption{The 1.76--4.2 keV images of the central region of Tycho's SNR with different smoothing radii: (a) smoothing radius $\sigma = 0$ pixel, (b) $\sigma = 2$ pixels and (c) $\sigma = 5$ pixels. The intensity is normalized by the total intensity and a pixel value of zero means the mean intensity. The blue and red contours show the intensity levels of $-1.5$ and $+1.5$, respectively.}
\label{fig:smoothimg}
\end{figure*}

Smoothing has an important and consequential effect on the genus curve of an image. On the other hand, observational images often have limited photon statistics, which we typically alleviate by smoothing. In this section we study the effect on the genus curve of applying different amounts of smoothing using a Gaussian filter in order to decide on an optimal balance between reducing noise fluctuations and smoothing away real image structures.


Figure \ref{fig:smoothimg} shows three images with different smoothing radii. In Figure \ref{fig:smoothimg} (a), the case with no smoothing, there are many small (noisy) contour levels that are mostly composed of isolated pixels. In particular, the regions in the blue contour ($\nu = -1.5$) have only $\sim$10 X-ray counts per pixel, leading to $\sim$30\% Poisson fluctuations in the pixel values.  These noise fluctuations introduce many spurious contour levels and bias the genus numbers. In the case of larger smoothing radii (Figure \ref{fig:smoothimg} b, c), the number of the contour levels from noise fluctuations are reduced significantly.

\begin{figure}[h]
 \begin{center}
  \includegraphics[bb=0 0 900 655,width=7cm]{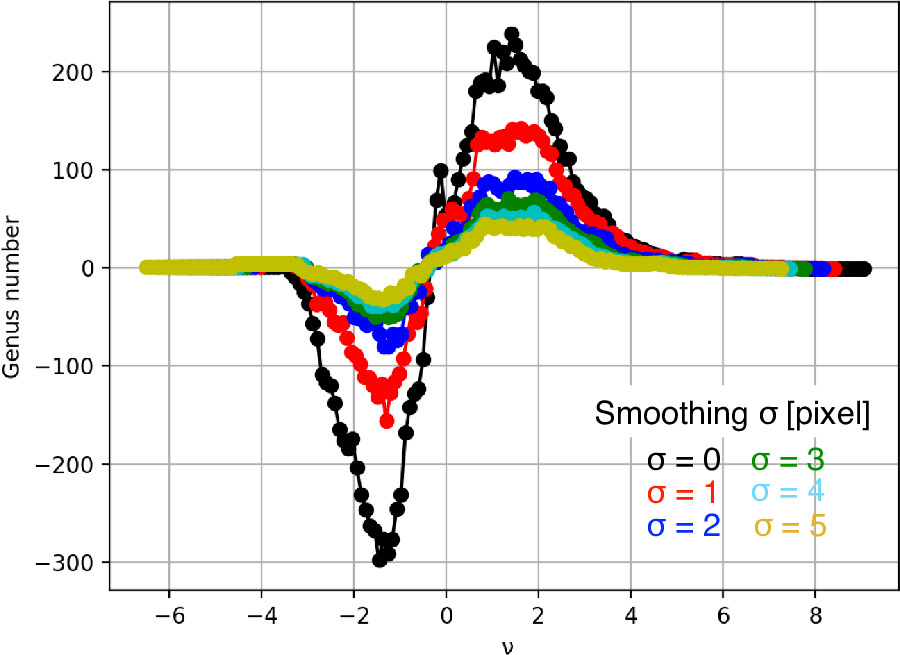}
 \end{center}
\caption{Genus curve for the central region of Tycho's SNR (1.76--4.2 keV) observed in 2009 with different smoothing $\sigma$. The color of each curve identifies the smoothing $\sigma$ used.}
\label{fig:GenusSmoothingNum}
\end{figure}

The genus curves obtained from the images in 2009 gradually change with smoothing radius (Figure \ref{fig:GenusSmoothingNum}).
In the case of no smoothing, the minimum of the genus numbers is about $-300$. We surmise that most of these ``holes" are due to Poisson fluctuations in the low count regime. For example, the absolute value of the minimum genus number is larger than that of the maximum genus number of $\sim$200, which means a hole-dominant morphology in Tycho's SNR. On the other hand, for smoothing with $\sigma>1$ pixel, the tendency is reversed (i.e., a clump-dominant morphology). In addition, the change of genus curve between the $\sigma =$ 0 and 1 pixel smoothing levels is the most significant (see also section \ref{sec:result}). The large change of the genus curve implies a transition from a noise dominant structure to a signal dominated one. Therefore, we believe that smoothing with a $\sigma$ of at least 1 pixel is necessary to obtain a reasonable genus curve without a large contribution of noisy pixels.

All the genus curves seem to have a skewed shape. For example, the genus curves do not pass through zero at the mean level (the intensity threshold $\nu = 0$ in Figure \ref{fig:GenusSmoothingNum}) and fall slowly at large threshold levels, which is similar to the genus curve for the chi-square pdf distribution (see red broken line in Figure \ref{fig:GaussGenuss}). \cite{2008ApJ...688.1021C} applied the genus statistic to the column density map of the Small Magellanic Cloud and evaluated whether the map was consistent with a Gaussian field using the value of the genus curve at the zero intensity point $\nu_0$. For a 3D field, positive and negative genus values at $\nu_0$ are classified into ``meatball" and ``swiss cheese" topologies, respectively. In the 2D case we have here, using a similar classification, the morphology of the clumpy structures in Tycho's SNR would be classified into peak-dominated because of the positive genus at $\nu_0$. In the following section, we carry out a more quantitative evaluation of the genus curves.

\begin{figure*}[h]
 \begin{center}
  \includegraphics[bb=0 0 1016 1094,width=16cm]{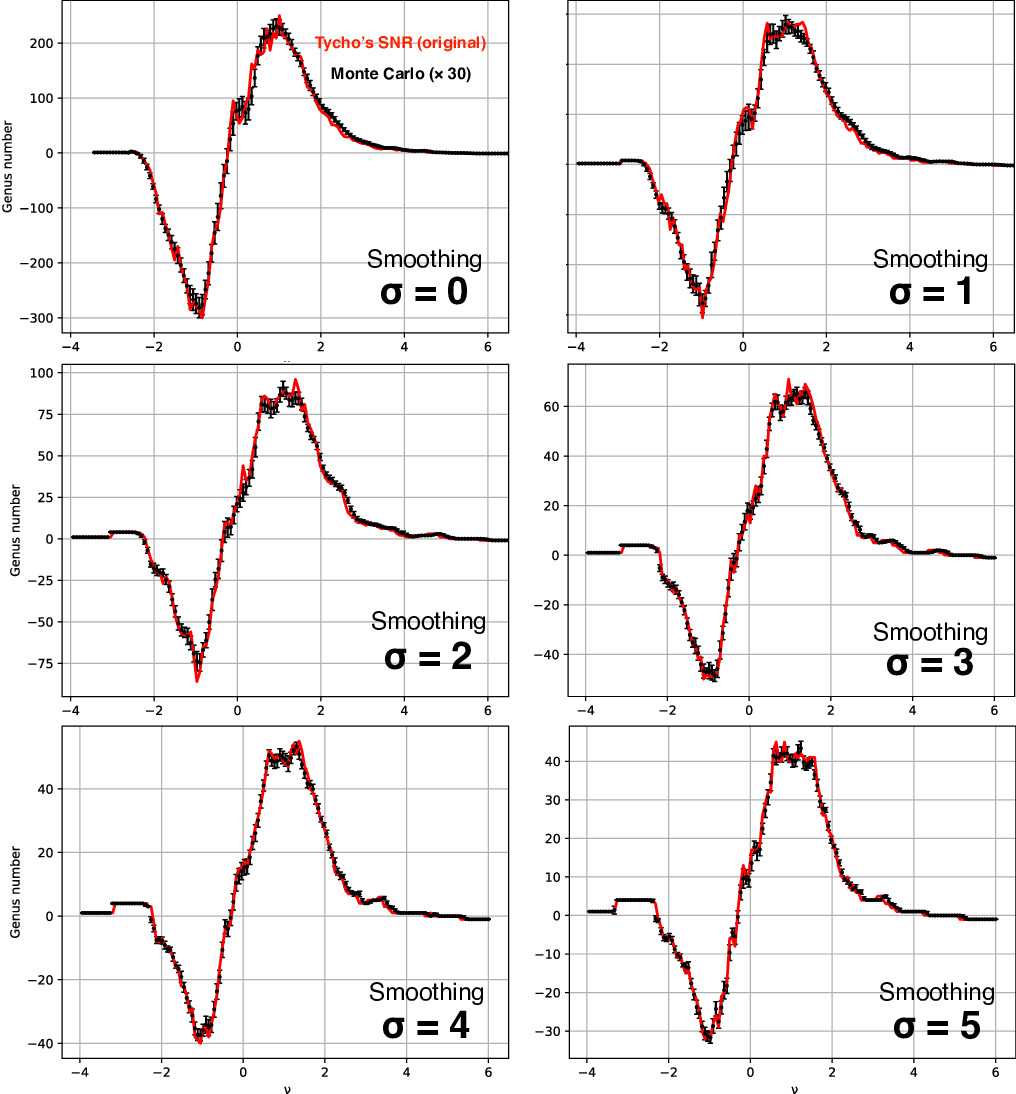}
 \end{center}
\caption{Genus curves with 1 $\sigma$ errors. The errors were estimated from Monte Carlo sampling. The red lines show the genus curve from the original image.}
\label{fig:Monte}
\end{figure*}

We generate uncertanties on the genus curves using a Monte Carlo resampling of the images. Here, we randomized the pixel values in the original X-ray image using a Poisson distribution and generated 30 sample images. As a result, we obtained 30 genus curves and estimated errors from the variance of the genus values (Figure \ref{fig:Monte}). We also determined the covariance matrix from these simulations and found little correlation between different bins in the genus curve.

\subsection{Genus Curves with Different Exposure-time Observations} \label{sec:exp}

\begin{figure}[h]
 \begin{center}
  \includegraphics[bb=0 0 649 1403,width=7cm]{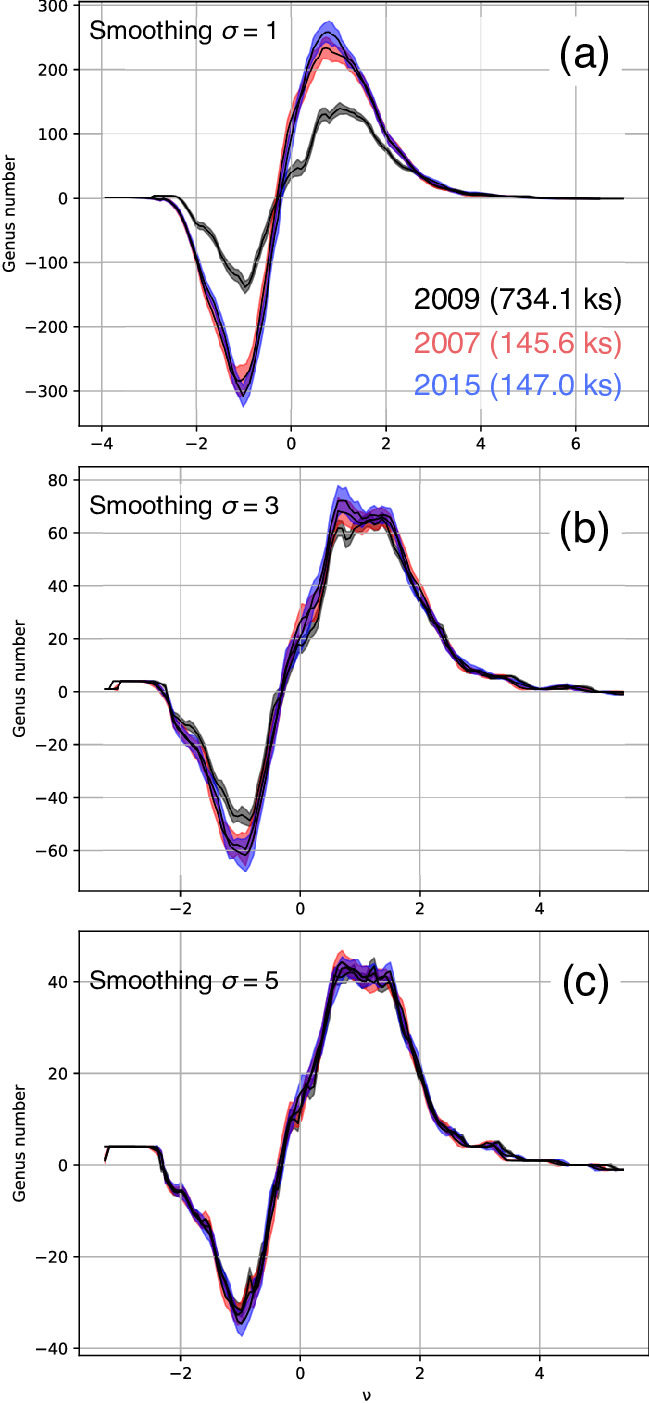}
 \end{center}
\caption{Genus curves for the central region of Tycho's SNR observed in 2007 (red), 2009 (black) and 2015 (green) using different smoothing amounts: (a) $\sigma$ = 1 pixels (= 0.492 arcsec), (b) $\sigma$ = 3 pixels (= 1.476 arcsec), and (c) $\sigma$ = 5 pixels (= 2.952 arcsec).}
\label{fig:GenusEpoch}
\end{figure}

As described in Section \ref{sec:smoothing}, noise from low-photon number statistics leads to an apparent bias in the genus curve, which can be reduced by image smoothing albeit at the cost of potentially removing some true image structure. Still it is difficult to determine the minimum smoothing radius required to obtain a reasonable genus curve using only a single observation. We here compare the genus curves among observations of different exposure length (Figure \ref{fig:GenusEpoch}).

In the case of $\sigma=1$ pixel smoothing, the observations in 2007 and 2015 show $\sim$1.5--2 times larger genus numbers than the $\sim$5 times longer observation in 2009 (Figure \ref{fig:GenusEpoch} a). This cleanly demonstrates the role of noise in biasing the genus curve. It is notable that the genus curves in 2007 and 2015 are similar to each other. Although Poisson fluctuations would result in different pixel values for each observation, the similar behavior of the genus curves for similar exposure times indicates that the average genus number count is a valid estimate across the ensemble of pixels in an image.

Figures \ref{fig:GenusEpoch} (b) and (c) indicate that large smoothing radii are effective in reducing the miscount of the genus number even for the short exposure time observations. Table \ref{tab:genusfit} shows a quantitative evaluation of the similarity among the genus curves using $\chi^2$ to compare the two short exposure datasets with the 5 times longer dataset for different amounts of smoothing. For smoothing with $\sigma>4$ pixels, we obtained nearly coincident genus curves among all the observations and statistical agreement between those curves.  This strongly supports that the genus curves are dominated by physical structures in the remnant, rather than noise, and that we can derive quantitative results from the curves (as we do in  \ref{sec:result} below).

\begin{table}[h]
\caption{The goodness of genus-curve fit between different observations$^\star$.}
\begin{center}
\begin{tabular}{cccccc}
\hline
  Smoothing $\sigma$                        &     $\chi^2$/d.o.f    & $\chi^2$/d.o.f \\
                                            &        2007--2009     &   2015--2009  \\ \hline
        0$^{\prime\prime}_{\cdot}$492  (1 pix)     &      2624.2/70        &  3316.8/70    \\
        0$^{\prime\prime}_{\cdot}$984  (2 pix)     &       638.6/70        &   639.7/70   \\
        1$^{\prime\prime}_{\cdot}$476  (3 pix)     &       172.5/70        &   204.9/70   \\
        1$^{\prime\prime}_{\cdot}$968  (4 pix)     &       72.5/70         &   82.2/70   \\
        2$^{\prime\prime}_{\cdot}$460  (5 pix)     &       75.5/70         &   77.4/70   \\\hline
\end{tabular}
\begin{tablenotes}
\item $^\star$ the fitting range is $-1.9 \lesssim \nu \lesssim +2.7$.
\end{tablenotes}
\label{tab:genusfit}
\end{center}
\end{table}

\subsection{Genus Curves with Different Energy Bands} \label{sec:energy}
In X-rays, we can investigate the distributions for each element using the energy band containing specific emission lines, which is very useful for understanding the distribution of elements associated with the explosion and their nucleosynthesis \citep[e.g.,][]{2000ApJ...528L.109H}. In particular, the clumpy ejecta model for Type Ia SNRs has been proposed to be caused by the $^{56}$Ni bubble effect \citep{2001ApJ...549.1119W}. If the $^{56}$Ni bubble made the ejecta clumps in Tycho's SNR, some evidence might be present in the elemental distributions. In this case, the comparison of the distributions between the Fe (because $^{56}$Ni decays to $^{56}$Fe) and the other elements would be important to understand the formation process of the clumps in the remnant.

\begin{figure}[h]
 \begin{center}
  \includegraphics[bb=0 0 543 682,width=7cm]{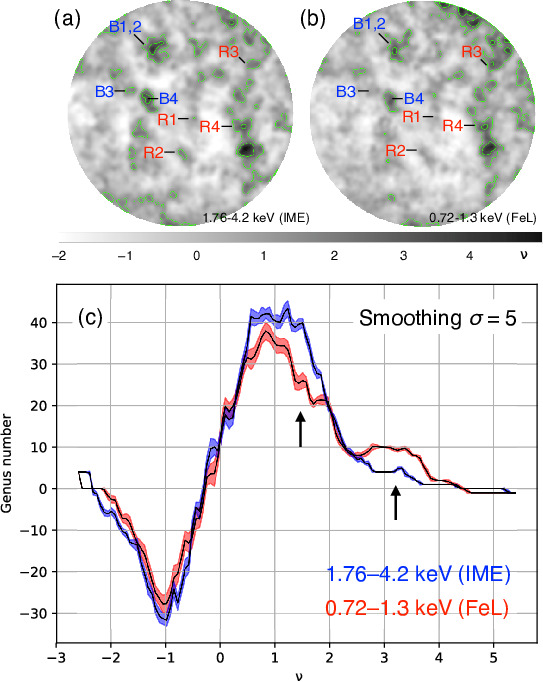}
 \end{center}
\caption{{\it Top:} X-ray images of the central region of the Tycho's SNR in (a) 1.76--4.2 keV and (b) 0.72--1.3 keV band with the smoothing $\sigma$ of 5 pixels. The green contours  show $\nu =$ 1.3 and 3.0. The 1.76--4.2 keV band includes the line emissions from the intermediate mass elements (IME) of Si, S, Ar and Ca. The 0.72--1.3 keV band includes the Fe-L shell emission lines. The ids of B1--4 and R1--4 correspond to the blueshifted and redshifted clumps, respectively, identified in \cite{2017ApJ...840..112S}. {\it Bottom:} (c) Genus curves of the 1.76--4.2 keV (black) and 0.72--1.3 keV (red) images with the smoothing $\sigma$ of 5 pixels.}
\label{fig:GenusIME-FeL}
\end{figure}

Figures \ref{fig:GenusIME-FeL} (a) and (b) show the comparison between the intermediate mass elements, IME (1.76--4.2 keV) and Fe (0.72--1.3 keV) images. Although the X-ray clumps in both images seem to show generally similar structures, there are some differences in the details. For example, the shapes of the blueshifted X-ray clumps B1 and B2 \citep[see also Figure 5 in][]{2017ApJ...840..112S} look different from each other, B4 has different number of intensity peaks, and the number of structures changes at the southeast and northwest side, etc. The differences between them are thought to appear in the genus curves (Figure \ref{fig:GenusIME-FeL} c) although the genus curve has no sensitivity to the detailed differences in shape. We here use the smoothing radius of 5 pixels to reduce the effect of the statistical uncertainty. Around $\nu = 1.3$, the genus number for the Fe image is $\sim$1.4 times smaller than that for the IME image, which seems to be mainly caused by the presence of the R1, R2 and southeastern clumps. On the other hand, around $\nu = 3$, the genus numbers for the Fe image become higher than that for the IME due to the presence of the northwestern clumps.  Effects of limb-brightening may be more apparent in the Fe image, since Fe emission is known to peak closer to the center of the remnant than the IME \citep{2005ApJ...634..376W}.

It is difficult to conclude that the difference of the distributions and their genus curves is caused by the $^{56}$Ni bubble effect using the current {\it Chandra} data. We use a broad energy-band image that has emission lines from other elements (e.g., Ne, Mg) as well as Fe because it is difficult to separate those emissions at ACIS-I spectral resolution. We could extract a pure Fe image from the Fe K$\alpha$ emission at $\sim$6.5 keV; however, the deepest available {\it Chandra} dataset has too few photons in this line for a useful comparison with the other elements. In addition, we have no simulation image to compare with our results. Deeper or larger effective area observations with high angular resolution (e.g., with {\it Chandra} and future missions like {\it Athena}, {\it Lynx}, {\it AXIS}) as well as hydrodynamical simulations including the $^{56}$Ni bubble and other relevant effects will be needed for more precise evaluation of the relative differences of the element distributions and the formation of clumps. 

\subsection{Evaluation of the Genus Curve of Tycho's SNR} \label{sec:result}

In order to evaluate the clumpy structure quantitatively, we fit the genus curve obtained from Tycho's SNR using the Gaussian and chi-square distribution analytic models (see Eq.~\ref{GaussEq} and \ref{chiEq}). As shown in section \ref{sec:smoothing}, Tycho's genus curve has a skewed shape, which implies a non-Gaussian distribution of clumps in the remnant. We can evaluate which model is more reasonable for the distribution of clumps by comparing the model fittings.

\begin{figure}[h]
 \begin{center}
  \includegraphics[bb=0 0 793 756,width=8cm]{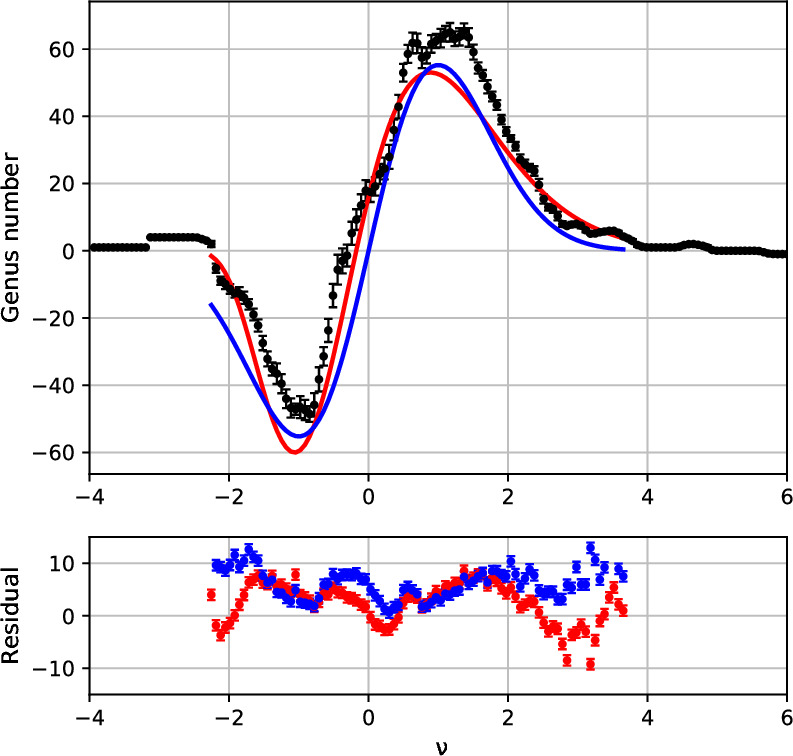}
 \end{center}
\caption{Genus curve for the central region of Tycho's SNR (smoothing with $\sigma = 3$ pixels equivalent to $\sigma = 1.476^{\prime\prime}$) and the best-fit analytic models. The blue and red lines show the analytic models for the Gaussian and chi-square pdf distribution, respectively.}
\label{fig:GenusFit}
\end{figure}

\begin{table}[h]
\scriptsize
\caption{The best-fit parameters of the genus curves obtained from Tycho's SNR using analytic models.}
\begin{center}
\begin{tabular}{cccccc}
\hline
&\multicolumn{2}{c}{\bf Gaussian random field} & \multicolumn{3}{c}{\bf chi-square random field}\\ 
    Smoothing $\sigma$        &   $\theta_{c}^\dagger$ &  $\chi^2$/d.o.f & $\theta_{c}^\dagger$    & $\chi^2$/d.o.f & $n^\star$\\ \hline
    0$^{\prime\prime}_{\cdot}$000      &       3.74$\pm$0.09        &       1064.6/84                &    3.76$\pm$0.05            &   396.2/83      &  52\\ 
        0$^{\prime\prime}_{\cdot}$492      &       4.91$^{+0.21}_{-0.24}$        &       2941.1/78                 &    4.61$\pm$0.07            &   353.4/78      &  19\\  
        0$^{\prime\prime}_{\cdot}$984      &     6$^{\prime\prime}_{\cdot}$26$^{+0.27}_{-0.31}$ &       3495.1/77                 &    6$^{\prime\prime}_{\cdot}$03$^{+0.12}_{-0.13}$    &   878.1/76      &  15\\  
        1$^{\prime\prime}_{\cdot}$476      &    7$^{\prime\prime}_{\cdot}$25$^{+0.29}_{-0.33}$      &   5226.7/87 &     7$^{\prime\prime}_{\cdot}$23$^{+0.16}_{-0.17}$ &       1813.3/88                  &  16\\
        1$^{\prime\prime}_{\cdot}$968      &    8$^{\prime\prime}_{\cdot}$26$^{+0.30}_{-0.34}$    &   4925.4/77 &     8$^{\prime\prime}_{\cdot}$38$^{+0.24}_{-0.26}$  &      2687.9/78                 &  26\\
        2$^{\prime\prime}_{\cdot}$460      &   8$^{\prime\prime}_{\cdot}$87$^{+0.28}_{-0.31}$  &   2541.2/63 &     9$^{\prime\prime}_{\cdot}$07$^{+0.25}_{-0.28}$  &      2076.8/64                 &  23\\ \hline
\end{tabular}
\begin{tablenotes}
\item $^\dagger$ the coherence angle of the genus curves estimated from the best-fit analytic models. The errors are $\Delta \chi^2 = 2.7 \times \chi^2_{\rm min}/{\rm d.o.f}$ level.
\item $^\star$ the order of the chi-square pdf distribution in the analytic model
\end{tablenotes}
\label{tab:fit}
\end{center}
\end{table}


 The fitting results are shown in Figure \ref{fig:GenusFit} and Table \ref{tab:fit}. In the fitting, only the coherence angle of the fluctuations, $\theta_c$, and the order of the chi-square distribution, $n$, are variables. The reduced $\chi^2$ values from the best fits indicate that the non-Gaussian, chi-square pdf, distribution is a more reasonable fit to explain the clumpy structure in the remnant (see $\chi^2$/d.o.f columns in Table \ref{tab:fit}), although the fits are not acceptable. In particular, at the minimum and peak of the genus curve, the genus values are  both numerically larger than those of the analytic models. This tendency is related to the peak-dominated structure of the ejecta distribution in the remnant.

The smoothing radius changes the fitting parameters (Table \ref{tab:fit}). In particular, the coherence angle  gradually increases as the smoothing radius increases in a fairly straightforward manner.
The coherence angle of fluctuations is also familiar in the studies of the microwave background radiation. We found the clumpy structures in Tycho's SNR show $\theta_c \sim$ 4--9$^{\prime\prime}$, which corresponds to $\sim$0.08--0.18 pc at the distance of 4 kpc. The best-fit values of the order of the chi-square pdf show similar values of $n =$ 15--26, except for the no smoothing case ($n =$ 52), which is noise dominated. 


\section{Discussion} \label{sec:sim}

We have applied genus statistics for the first time to Tycho's supernova remnant (SN 1572), to quantify the morphology of its clumpy structures. We found a skewed shape for the genus curve, which indicates that the non-Gaussian (e.g., chi-square) distribution of clumps is more favorable to explain the structures in the remnant rather than the Gaussian distribution (although neither of these models provides a statistically acceptable fit). Our results also showed subtle differences in the genus curves for the images generated by emission from iron and IMEs. In this section we examine the genus curves to help  understand the formation of clumps in Type Ia SNRs.

\begin{figure*}[t!]
 \begin{center}
  \includegraphics[bb=0 0 630 440,width=17cm]{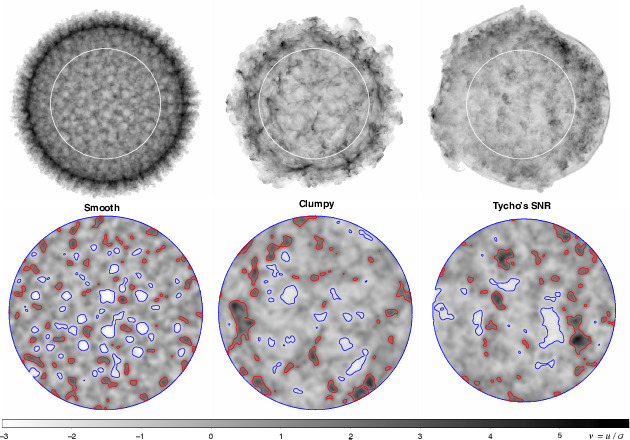}
 \end{center}
\caption{{\it Top:} the sqrt scale images of the smooth ejecta model (left), the clumpy ejecta model (middle) and Tycho's SNR. Only the image of Tycho's SNR is smoothed by the smoothing $\sigma$ = 1.5 pixels. {\it Bottom:} the images of the central region of the smooth ejecta model (left), the clumpy ejecta model (middle) and Tycho's SNR with the smoothing $\sigma$ = 5 pixels. The blue and red contours show $\nu$ = $-1.5$ and $+1.5$, respectively.}
\label{fig:Models}
\end{figure*}

\cite{2017ApJ...842...28W} used a simple kinematic quantity (the distribution of deceleration parameters for isolated clumps) to compare observations of Tycho's SNR with two hydrodynamic simulations evolved from near the time of the initial explosion to the current age of the remnant. The two models correspond to two different cases for the initial ejecta distribution: one smooth and the other clumpy.  These authors found that both simulations produced a range of deceleration parameters that were in agreement with observations. In spite of this apparent consistency with the observations, the simulated images showed distinctly different visual appearances in their ejecta clump distributions (Figure \ref{fig:Models}). 
Here we use the genus statistic for a quantitative comparison between the observations and simulations.

\begin{table}[h]
\caption{Summary of the best-fit parameters of the genus curves.}
\begin{center}
\begin{tabular}{cccccc}
\hline&\multicolumn{2}{c}{\bf Gaussian pdf} & \multicolumn{3}{c}{\bf chi-square pdf}\\

Smoothing $\sigma$       &   $d_{\rm Gauss}^{\ast}$& $\theta_{c}^\dagger$    &  $d_{\chi^2_n}^{\ast}$ & $\theta_{c}^\dagger$ & $n^\star$\\ \hline
{\bf Smooth model} &&&&\\
        0$^{\prime\prime}_{\cdot}$8 (= 2 pix)      &       7.9         &    4$^{\prime\prime}_{\cdot}$9            &       7.4                 &   4$^{\prime\prime}_{\cdot}$8      &  $\geq$271\\  
        1$^{\prime\prime}_{\cdot}$6 (= 4 pix)      &       6.4         &    5$^{\prime\prime}_{\cdot}$7            &       6.1                 &   5$^{\prime\prime}_{\cdot}$7      &  221\\
        2$^{\prime\prime}_{\cdot}$4 (= 6 pix)      &       4.7         &    6$^{\prime\prime}_{\cdot}$4            &       4.8                 &   6$^{\prime\prime}_{\cdot}$3      &  $\geq$269\\ \hline
{\bf Clumpy model} &&&&\\
        0$^{\prime\prime}_{\cdot}$8 (= 2 pix)      &       8.7         &    6$^{\prime\prime}_{\cdot}$6            &       4.6                 &   6$^{\prime\prime}_{\cdot}$5      &  17\\  
        1$^{\prime\prime}_{\cdot}$6 (= 4 pix)      &       7.2         &    7$^{\prime\prime}_{\cdot}$9            &       3.9                 &   7$^{\prime\prime}_{\cdot}$8      &  15\\
        2$^{\prime\prime}_{\cdot}$4 (= 6 pix)      &       6.6         &    9$^{\prime\prime}_{\cdot}$2            &       3.7                 &   9$^{\prime\prime}_{\cdot}$0      &  12\\ \hline
\end{tabular}
\begin{tablenotes}
\item $^\ast$ the mean genus distance from the analytic models.
\item $^\dagger$ the coherence angle of the genus curves estimated from the best-fit analytic models.
\item $^\star$ the order of the random chi-square pdf distribution.
\end{tablenotes}
\label{tab:fit3}
\end{center}
\end{table}

Figure \ref{fig:Models} compares the images among the smooth and clumpy ejecta models \citep[both from][]{2017ApJ...842...28W} and the observations of Tycho's SNR. The pixel size in the model images is scaled by the radius of the remnant, which corresponds to an angular size of 0.4 arcsec. In the case of the smooth ejecta model, clumps arise from the action of the dynamical instabilities acting through the duration of the remnant's evolution. Here the structures appear to be homogeneously distributed across the central region and the clumps are of similar shape and size.  On the other hand, the clumps in the clumpy ejecta model appear more filamentary and are distributed more randomly. Both models assume the an exponential radial density profile of \cite{1998ApJ...497..807D} and explosion parameters of 10$^{51}$ erg with 1.4 $M_{\odot}$ of ejecta. The clumpy model is produced using a Perlin algorithm to generate noise with a maximum angular scale of $\sim$20$^{\circ}$ and a maximum-to-minimum density contrast of 6. These simulations using the hydrodynamics code VH-1 are described in detail in \cite{2013MNRAS.429.3099W}. 

\begin{figure}[h]
 \begin{center}
  \includegraphics[bb=0 0 762 1108,width=7cm]{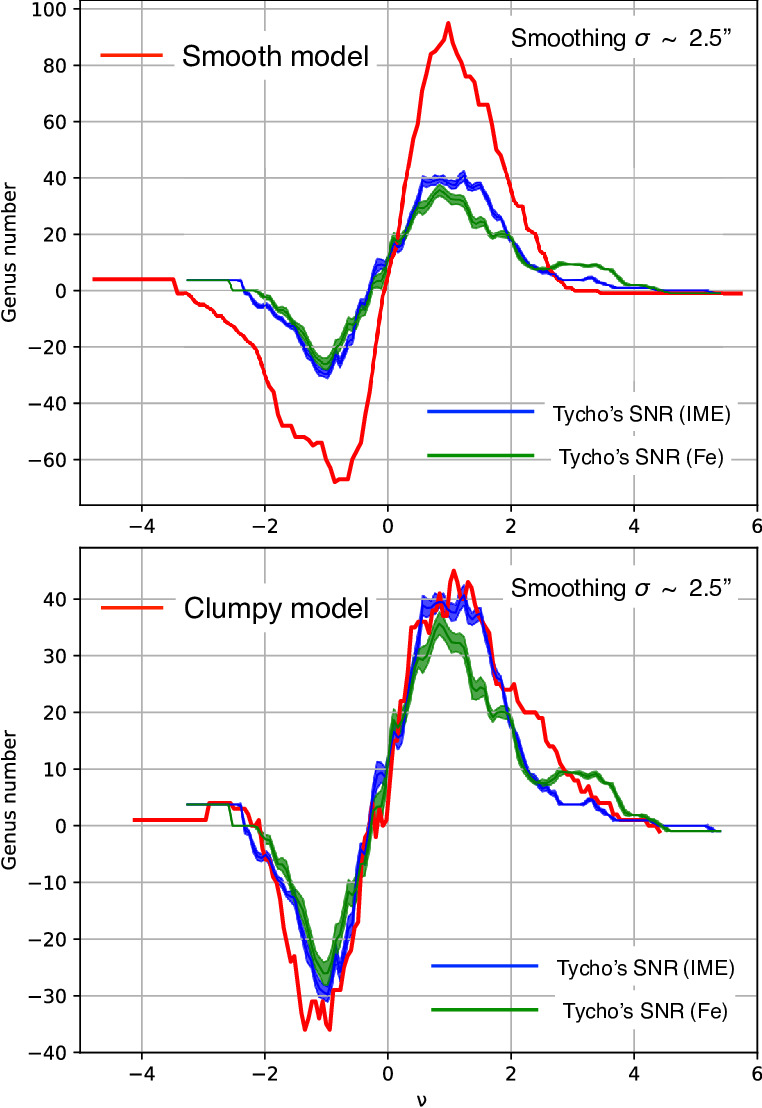}
 \end{center}
\caption{Comparison of the genus curves (in red) for the smooth (top) and clumped initial ejecta (bottom) models  for an image smoothing of $\sigma$ = 6 pixels (= 2$^{\prime\prime}_{\cdot}$4). 
The blue (green) lines show the genus curves for Tycho's SNR with an image smoothing $\sigma$ = 5 pixels (= 2$^{\prime\prime}_{\cdot}$46) for the energy band dominated by IME (Fe) emission.}
\label{fig:GenusComp}
\end{figure}

Figure \ref{fig:GenusComp} and Table \ref{tab:fit3} show the comparison of the genus curves among the smooth ejecta model, the clumpy ejecta model and Tycho's SNR.  Since we do not have uncertainties on the model genus curves we cannot use $\chi^2$ as the figure-of-merit function; instead for the comparison here we use the mean genus distance 
\begin{equation}
        d_{\rm genus} = \frac{1}{n}\sum^n_i  | G_{\rm data}(I_i) - G_{\rm model}(I_i) |,
\end{equation}
where $G(I_i)$ is the genus number at the intensity threshold $I_i$. We define the best-fit as the model with the minimum genus distance. 


The fits to the genus curve of the smooth model for both the Gaussian and chi-square distributions yield very similar coherence angles and $d_{\rm genus}$ values.  Additionally the chi-square distribution requires large $n$ values ($>$200) for the best fit which is securely in the regime where the chi-square random field approaches a Gaussian random field. Thus we conclude that the distribution of clumps in the smooth model is close to a random Gaussian distribution. The genus numbers for the smooth model have much larger absolute values than those of the genus curve for Tycho's SNR at any smoothing $\sigma$ (for one specific smoothing level see Fig.~\ref{fig:GenusComp}).

The clumpy model also shows similar values for the coherence angles between the Gaussian and chi-square distributions, but in this case the genus distance is considerably less for the chi-square distribution, implying a better description of the genus curve. This is similar to what we found for the observation of Tycho's SNR (see section \ref{sec:result}).  Additionally the coherence angles from the clumpy model are a better match to the data for all smoothing scales. And the genus curve for the clumpy model is similar to that of the observed remnant, in particular at a smoothing of $\sigma$ = 5 pixels (Figure \ref{fig:GenusComp}). The genus statistic therefore strongly supports an initial clumped ejecta distribution as the origin of the clumps in Tycho's supernova remnant.

\cite{2013MNRAS.429.3099W} argued that the presence of ejecta knots ahead of the forward shock in Tycho's SNR and SN 1006 can be generated by smooth ejecta without any initial clumpiness using their three-dimensional hydrodynamics simulations. In order to approximate the effect of efficient particle acceleration, they allowed for the adiabatic index $\gamma$ to be a model parameter.  The simulations were able to produce clumps breaking through the mean shock radius as observed but only if the shock compression was quite high (a compression ratio of 11). Also, they found that ignoring the emission from material below a certain ionization age tended to change the observed morphology. In fact, the diversity in imaging introduced by changing the adiabatic index and the ionization age cut-offs are not considered in the smooth model we used. Those effects might change the genus curve for the smooth ejecta model. Additionally, we expect differences in the genus curves for each model from different random realizations of the initial conditions.  Viewing along different lines-of-sight would also produce different genus curves. Investigation of such studies will be the focus of future work.

The existence of a high velocity feature (HVF) identified as the Ca II triplet at a velocity of 20,000--24,000 km s$^{-1}$ in the light-echo spectrum of SN 1572 \citep{2008Natur.456..617K} offers support for initial clumping in the ejecta. Similar HVFs have been found in many SNe \citep[e.g.,][]{2005ApJ...623L..37M}, as a result of asphericity in the explosion due to, for example, accretion from a companion or an intrinsic effect of the explosion itself \citep[e.g.,][]{2003ApJ...591.1110W,2003ApJ...593..788K,2006ApJ...645..470T}. \cite{2003ApJ...593..788K} analyzed both spectroscopy and spectropolarimetry of SN 2001el and showed that both an aspherical photosphere and a single high-velocity blob can reproduce the observations. Also three-dimensional models suggest that large blobs (opening angle: $\sim$80$^{\circ}$) or a thick torus (opening angle: $\sim$60$^{\circ}$) can naturally explain the observed diversity in strength of HVFs \citep[][]{2006ApJ...645..470T} although these structures would be much larger in physical size than the X-ray clumps in Tycho's SNR. At a minimum, the existence of a HVF implies some kind of clumpiness in the ejecta of SN 1572.

The theoretical  origin of the initial clumpiness in SN Ia ejecta is still unclear.
Important information for the initial ejecta clumping might be in the ignition mechanism.
For example, the ignition of the thermonuclear deflagration flame at the end of the convective carbon ``simmering" phase in the core of the white dwarf is still not well understood, and much about the ignition kernel distribution remains unknown.
\cite{2005A&A...430..585G} simulated a pure deflagration model with multiple ignition points, which provided 4--5 large clumps of $^{56}$Ni at the photosphere at the time of maximum brightness. Large $^{56}$Ni clumps seem to be a common feature in all 3D pure deflagration models, however this is in some conflict with the remarkable homogeneity of the observed SNe Ia sample \citep{2002ApJ...567.1037T}. Also we do not find such large structures in SN Ia remnants. On the other hand, a delayed detonation model could potentially alleviate this problem by breaking the clumps when the detonation hits them.
In fact such large $^{56}$Ni clumps seen in the pure deflagration model do not appear in the three-dimensional delayed-detonation models with multiple ignition points  \citep[e.g.,][]{2008A&A...478..843B,2011MNRAS.414.2709S,2013MNRAS.429.1156S}. 


Burning in an aspherical manner as seen in ``gravitational confined detonations" \citep[e.g.,][]{2004ApJ...612L..37P,2008ApJ...681.1448J,2007ApJ...668.1132R} also could break the smooth ejecta distribution hypothesis. In this mechanism, ignition occurs at one or several off-center points, which makes a burning bubble of buoyant hot ash that ascends rapidly. Rising out of the star, it breaks through the surface, and a shock is driven around the surface which collides at the point opposite the breakout. This model creates a large blob of ash in the upper hemisphere of the white dwarf. However, it produces large amounts of $^{56}$Ni and small amounts of intermediate-mass elements, which implies a high-luminosity Type Ia supernovae \citep[][]{2008ApJ...681.1448J}. In addition, the element distributions are expected to be highly asymmetric \citep[see also ][]{2010ApJ...712..624M}. These characteristics are not consistent with the light echo spectrum of SN 1572, which appears to be a normal SN Ia \citep{2008Natur.456..617K}.


\begin{figure}[h]
 \begin{center}
  \includegraphics[bb=0 0 1097 825,width=8cm]{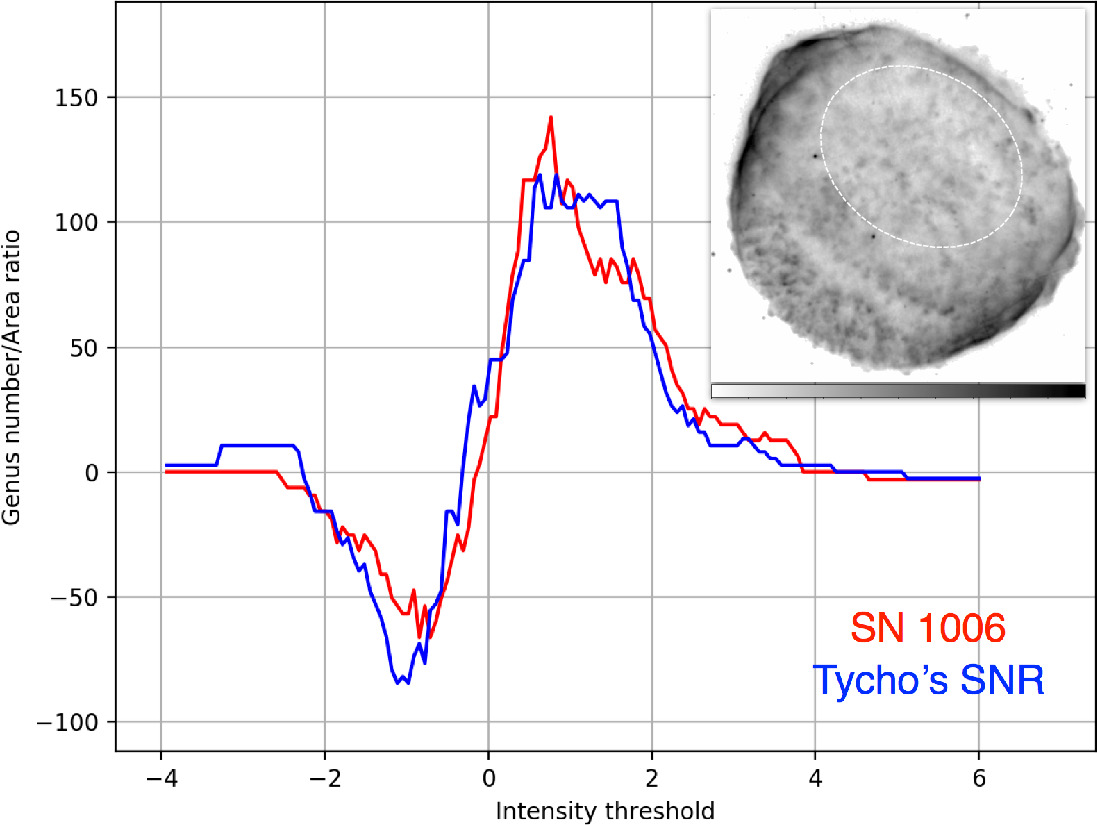}
 \end{center}
\caption{Comparison of the genus curves for SN 1006 (red) and Tycho's SNR (blue). The smoothing scales for obtaining the genus curves are similar to each other ($\sim$0.01 $\times$ radius). The genus numbers are normalized by the area ratio (the extracted area / the whole SNR area). The small inset figure at the upper right shows the {\it XMM-Newton} image (0.5--2 keV) with the overplotted ellipse showing the area used for generating the genus curve. The {\it XMM} image was made from the Large Program observations in 2008--2010 (PI: Anne Decourchelle).}
\label{fig:GenusSN1006}
\end{figure}

Applying the genus statistic to other Type Ia SNRs will allow us to test the variation in clump properties from one explosion to another. Perhaps the next best case after Tycho's SNR would be SN 1006 where arcminute-scale X-ray clumps have been observed \citep[e.g.,][]{2011ApJ...735L..21R,2014ApJ...781...65W}. 
%
%
In Figure \ref{fig:GenusSN1006} we present a first look at the genus curve of SN 1006 obtained from the  {\it XMM-Newton}  Large Program observations over 2008--2010 (PI: Anne Decourchelle). We selected a sub-region of the remnant away from obvious regions with clear limb-brightening.  SN 1006's genus curve closely resembles that of Tycho's SNR with a similar asymmetric and skewed shape. The implication  we draw from this preliminary comparison is that the ejecta distribution in SN 1006 is also unlikely to be explained by a smooth ejecta model. Another SN Ia remnant is Kepler's SNR where kinematically-young X-ray-emitting small-scale knots have been identified \citep{2017ApJ...845..167S}. However, applying the genus statistic to this remnant is more problematic because of the complex X-ray surface brightness distribution and kinematics of the X-ray-emitting ejecta \citep[e.g.,][]{2007ApJ...668L.135R,2018PASJ...70...88K}.   These results, however, do bear on the clumpy nature of the initial ejecta distribution in SNe Ia.

The variety of possible SN Ia explosion mechanisms and the difficulty and cost of running end-to-end 3D hydrodyamical simulations from realistic SN explosions to the SNR phase hundreds of years later make it unlikely that high fidelity models of SNR structure will be available anytime soon.  So investigation of the properties of the initial ejecta distribution  required to explain the structure of SNRs seems to us to be a useful area to research.  In the clumpy model we used here, the density contrast range was a factor of six.  Comparison of the genus curves for the clumpy model and the observations of Tycho's SNR (Fig.~\ref{fig:GenusComp}, bottom panel) shows that the model produces higher absolute genus numbers than the data at high and low count levels (i.e., around $\nu \sim 2.5$ and $\nu \sim -1.5$). Reducing the initial density contrast in the ejecta may reduce these discrepancies and provide a better model fit to the observations.  This information can then feed back  as constraints on the output of SN Ia explosion models.

\section{Summary} \label{sec:conclusion}

We have made the first application of the genus statistic to X-ray images of young remnants of SN Ia, specifically the remnant of SN 1572 (Tycho's SNR), in order to constrain the properties of clumps in the initial ejecta distribution. The genus curve of Tycho's remnant shows a skewed shape, which strongly indicates that the distribution of clumps follows a non-Gaussian distribution. Comparison of the observed genus curve with curves generated from two separate 3D hydrodynamical simulations of Tycho's SNR assuming  smooth and clumped initial ejecta distributions reveals a clear preference for the clumpy ejecta model.  This is strong evidence that SN Ia explosions produce ejecta that are  significantly clumped and that measurements of the properties of the clumps are possible from X-ray images.

At present, the cause of the initial ejecta clumping in SNe Ia is still theoretically unclear and end-to-end 3D simulations from  the SN explosion to the SNR are not yet available.  Running 3D simulations of parameterized  initial ejecta clumping distribution and comparing their genus curves to the image data will be an effective next step to help make progress on the nature of Type Ia SNe.

\acknowledgments{We are very grateful to Prof. John Blondin for kindly providing us the hydrodynamical models. T.S.\ was supported by the Japan Society for the Promotion of Science (JSPS) KAKENHI Grant Number JP18H05865 and the Special Postdoctoral Researchers Program in RIKEN. JPH acknowledges support for X-ray studies of SNRs from NASA grant NNX15AK71G to Rutgers University.
}




\begin{thebibliography}{}


\bibitem[Bravo \& Garc{\'{\i}}a-Senz(2008)]{2008A&A...478..843B} Bravo, E., \& Garc{\'{\i}}a-Senz, D.\ 2008, \aap, 478, 843

\bibitem[Chepurnov et al.(2008)]{2008ApJ...688.1021C} Chepurnov, A., Gordon, J., Lazarian, A., \& Stanimirovic, S.\ 2008, \apj, 688, 1021


\bibitem[Coles(1988)]{1988MNRAS.234..509C} Coles, P.\ 1988, \mnras, 234, 509 

\bibitem[Decourchelle, Ellison, 
\& Ballet(2000)]{2000ApJ...543L..57D} Decourchelle, A., Ellison, D.~C., \& Ballet, J.\ 2000, \apjl, 543, L57 

\bibitem[Decourchelle et al.(2001)]{2001A&A...365L.218D} Decourchelle, A., Sauvageot, J.~L., Audard, M., et al.\ 2001, \aap, 365, L218 

\bibitem[Dwarkadas \& Chevalier(1998)]{1998ApJ...497..807D} Dwarkadas, V.~V., \& Chevalier, R.~A.\ 1998, \apj, 497, 807 

\bibitem[Garc{\'{\i}}a-Senz \& Bravo(2005)]{2005A&A...430..585G} Garc{\'{\i}}a-Senz, D., \& Bravo, E.\ 2005, \aap, 430, 585 

\bibitem[Gott et al.(1986)]{1986ApJ...306..341G} Gott, J.~R., III, Melott, A.~L., \& Dickinson, M.\ 1986, \apj, 306, 341 

\bibitem[Gott et al.(1987)]{1987ApJ...319....1G} Gott, J.~R., III, Weinberg, D.~H., \& Melott, A.~L.\ 1987, \apj, 319, 1 

\bibitem[Gull(1975)]{1975MNRAS.171..263G} Gull, S.~F.\ 1975, \mnras, 171, 263 

\bibitem[Hughes(2000)]{2000ApJ...545L..53H} Hughes, J.~P.\ 2000, \apjl, 545, L53 

\bibitem[Hughes et al.(2000)]{2000ApJ...528L.109H} Hughes, J.~P., Rakowski, C.~E., Burrows, D.~N., \& Slane, P.~O.\ 2000, \apjl, 528, L109 

\bibitem[Hwang \& Gotthelf(1997)]{1997ApJ...475..665H} Hwang, U., \& Gotthelf, E.~V.\ 1997, \apj, 475, 665 

\bibitem[Hwang et al.(2002)]{2002ApJ...581.1101H} Hwang, U., Decourchelle, A., Holt, S.~S., \& Petre, R.\ 2002, \apj, 581, 1101

\bibitem[Hwang et al.(2000)]{2000ApJ...537L.119H} Hwang, U., Holt, S.~S., \& Petre, R.\ 2000, \apjl, 537, L119 


\bibitem[Jordan et al.(2008)]{2008ApJ...681.1448J} Jordan, G.~C., IV, Fisher, R.~T., Townsley, D.~M., et al.\ 2008, \apj, 681, 1448

\bibitem[Koch et al.(2017)]{2017MNRAS.471.1506K} Koch, E.~W., Ward, C.~G., Offner, S., Loeppky, J.~L., \& Rosolowsky, E.~W.\ 2017, \mnras, 471, 1506 

\bibitem[Kasen et al.(2003)]{2003ApJ...593..788K} Kasen, D., Nugent, P., Wang, L., et al.\ 2003, \apj, 593, 788 

\bibitem[Kasuga et al.(2018)]{2018PASJ...70...88K} Kasuga, T., Sato, T., Mori, K., Yamaguchi, H., \& Bamba, A.\ 2018, \pasj, 70, 88 

\bibitem[Kowal et al.(2007)]{2007ApJ...658..423K} Kowal, G., Lazarian, A., \& Beresnyak, A.\ 2007, \apj, 658, 423 

\bibitem[Krause et al.(2008)]{2008Natur.456..617K} Krause, O., Tanaka, M., Usuda, T., et al.\ 2008, \nat, 456, 617

\bibitem[Lopez et al.(2009)]{2009ApJ...691..875L} Lopez, L.~A., Ramirez-Ruiz, E., Pooley, D.~A., \& Jeltema, T.~E.\ 2009, \apj, 691, 875 

\bibitem[Maeda et al.(2010)]{2010ApJ...712..624M} Maeda, K., R{\"o}pke, F.~K., Fink, M., et al.\ 2010, \apj, 712, 624

\bibitem[Mazzali et al.(2005)]{2005ApJ...623L..37M} Mazzali, P.~A., Benetti, S., Altavilla, G., et al.\ 2005, \apjl, 623, L37 

\bibitem[Melott et al.(1989)]{1989ApJ...345..618M} Melott, A.~L., Cohen, A.~P., Hamilton, A.~J.~S., Gott, J.~R., III, \& Weinberg, D.~H.\ 1989, \apj, 345, 618 

\bibitem[Orlando et al.(2012)]{2012ApJ...749..156O} Orlando, S., Bocchino, F., Miceli, M., Petruk, O., \& Pumo, M.~L.\ 2012, \apj, 749, 156 

\bibitem[Perlmutter et al.(1999)]{1999ApJ...517..565P} Perlmutter, S., Aldering, G., Goldhaber, G., et al.\ 1999, \apj, 517, 565 

\bibitem[Plewa et al.(2004)]{2004ApJ...612L..37P} Plewa, T., Calder, A.~C., \& Lamb, D.~Q.\ 2004, \apjl, 612, L37

\bibitem[Rakowski et al.(2011)]{2011ApJ...735L..21R} Rakowski, C.~E., Laming, J.~M., Hwang, U., et al.\ 2011, \apjl, 735, L21 

\bibitem[Reynolds et al.(2007)]{2007ApJ...668L.135R} Reynolds, S.~P., Borkowski, K.~J., Hwang, U., et al.\ 2007, \apjl, 668, L135 

\bibitem[Riess et al.(1998)]{1998AJ....116.1009R} Riess, A.~G., Filippenko, A.~V., Challis, P., et al.\ 1998, \aj, 116, 1009 

\bibitem[R{\"o}pke et al.(2007)]{2007ApJ...668.1132R} R{\"o}pke, F.~K., Hillebrandt, W., Schmidt, W., et al.\ 2007, \apj, 668, 1132

\bibitem[Sato \& Hughes(2017a)]{2017ApJ...840..112S} Sato, T., \& Hughes, J.~P.\ 2017a, \apj, 840, 112 

\bibitem[Sato \& Hughes(2017b)]{2017ApJ...845..167S} Sato, T., \& Hughes, J.~P.\ 2017b, \apj, 845, 167

\bibitem[Seitenzahl et al.(2011)]{2011MNRAS.414.2709S} Seitenzahl, I.~R., Ciaraldi-Schoolmann, F., \& R{\"o}pke, F.~K.\ 2011, \mnras, 414, 2709

\bibitem[Seitenzahl et al.(2013)]{2013MNRAS.429.1156S} Seitenzahl, I.~R., Ciaraldi-Schoolmann, F., R{\"o}pke, F.~K., et al.\ 2013, \mnras, 429, 1156

\bibitem[Seward et al.(1983)]{1983ApJ...266..287S} Seward, F., Gorenstein, P., \& Tucker, W.\ 1983, \apj, 266, 287 

\bibitem[Tanaka et al.(2006)]{2006ApJ...645..470T} Tanaka, M., Mazzali, P.~A., Maeda, K., \& Nomoto, K.\ 2006, \apj, 645, 470 

\bibitem[Thomas et al.(2002)]{2002ApJ...567.1037T} Thomas, R.~C., Kasen, D., Branch, D., \& Baron, E.\ 2002, \apj, 567, 1037

\bibitem[Vancura et al.(1995)]{1995ApJ...441..680V} Vancura, O., Gorenstein, P., \& Hughes, J.~P.\ 1995, \apj, 441, 680 

\bibitem[Wang \& Chevalier(2001)]{2001ApJ...549.1119W} Wang, C.-Y., \& Chevalier, R.~A.\ 2001, \apj, 549, 1119

\bibitem[Wang et al.(2003)]{2003ApJ...591.1110W} Wang, L., Baade, D., H{\"o}flich, P., et al.\ 2003, \apj, 591, 1110 

\bibitem[Warren \& Blondin(2013)]{2013MNRAS.429.3099W} Warren, D.~C., \& Blondin, J.~M.\ 2013, \mnras, 429, 3099 

\bibitem[Warren et al.(2005)]{2005ApJ...634..376W} Warren, J.~S., Hughes, J.~P., Badenes, C., et al.\ 2005, \apj, 634, 376 

\bibitem[Williams et al.(2017)]{2017ApJ...842...28W} Williams, B.~J., Coyle, N.~M., Yamaguchi, H., et al.\ 2017, \apj, 842, 28

\bibitem[Winkler et al.(2014)]{2014ApJ...781...65W} Winkler, P.~F., Williams, B.~J., Reynolds, S.~P., et al.\ 2014, \apj, 781, 65

\bibitem[Yamaguchi et al.(2017)]{2017ApJ...834..124Y} Yamaguchi, H., Hughes, J.~P., Badenes, C., et al.\ 2017, \apj, 834, 124


\end{thebibliography}
\end{document}